\documentclass[%
 aip,
 sd,%
 amsmath,amssymb,
 reprint,%
 longbibliography
]{revtex4-1}

\usepackage{graphicx,color}
\usepackage{dcolumn}
\usepackage{amsmath}
\usepackage{amssymb}
\usepackage{epstopdf}

\usepackage[]{hyperref}
\hypersetup{colorlinks=true,linkcolor=blue,citecolor=blue,urlcolor=blue,pdfpagemode=UseNone}

\begin{document}


\title{Optically detected magnetic resonance of nitrogen vacancies in a diamond anvil cell using designer diamond anvils}

\author{L. Steele}
\thanks{These authors contributed equally to this work.}
\author{M. Lawson}
\thanks{These authors contributed equally to this work.}
\affiliation{Department of Physics, University of California, Davis, California 95616 USA.}
\author{M. Onyszczak}
\affiliation{Department of Physics and Astronomy, Iowa State University, Ames, Iowa 50011, USA}
\author{B. T. Bush}
\author{Z. Mei}
\affiliation{Department of Physics, University of California, Davis, California 95616 USA.}
\author{A. P. Dioguardi}
\affiliation{Los Alamos National Laboratory, Los Alamos, New Mexico 87545, USA}
\author{J. King}
\author{A. Parker}
\author{A. Pines}
\affiliation{Department of Chemistry, University of California, Berkeley, California 94720, USA, and
Materials Sciences Division, Lawrence Berkeley National Laboratory, Berkeley, California 94720, USA.}
\author{S. T. Weir}
\author{W.  Evans}
\author{K. Visbeck}
\affiliation{Lawrence Livermore National Laboratory, University of California, P.O. Box 808, Livermore,
California 94550}
\author{Y. K. Vohra}
\affiliation{Department of Physics, University of Alabama at Birmingham, Birmingham, Alabama 35294-1170, USA.}
\author{N.J. Curro}
\affiliation{Department of Physics, University of California, Davis, California 95616 USA.}

\date{\today}

\begin{abstract}
Optically detected magnetic resonance of nitrogen vacancy centers in diamond offers novel routes to both DC and AC magnetometry in diamond anvil cells under high pressures ($>3$ GPa).  However, a serious challenge to realizing experiments has been the insertion of microwave radiation in to the sample space without screening by the gasket material.  We utilize designer anvils with lithographically-deposited metallic microchannels on the diamond culet as a microwave antenna.  We detected the spin resonance of an ensemble of microdiamonds under pressure, and measure the pressure dependence of the zero field splitting parameters.  These experiments enable the possibility for all-optical magnetic resonance experiments on sub-$\mu$L sample volumes at high pressures.
\end{abstract}

\maketitle

\section{\label{sec:level1}INTRODUCTION}

A number of materials exhibit dramatic changes in electronic behavior at high pressure, and there is a great need to develop experimental tools to investigate electronic matter under extreme conditions. \cite{HeggerRh115discovery,Cambridge122pressure2009,Dias715}  Magnetic resonance is an important tool that can be realized under pressure and can provide important microscopic information about the electronic degrees of freedom.\cite{TakigawaSr122pressure,Lin2015}  Although bulk measurements have been realized at pressures up to 500 GPa,\cite{Dias715} magnetic resonance has been limited to pressures below 20 GPa.\cite{KitagawaPressureCell,HaaseJMR2015}  Pressures greater than 4 GPa can usually only be achieved by anvil cells, which present significant technical challenges because not only is the sample space limited (typically on the order of 10-100 nL), but an inductive coil must also surround the sample in order to excite and detect the resonance.  In some cases, this coil can be located outside of the gasket,\cite{SilveraNMR} but in such a case the signal to noise ratio is reduced due to poor filling fraction.  Another approach is to locate the coil inside the gasket, requiring insulated leads to pass between the diamond culet and gasket material without severing under high pressure.\cite{HaaseJMR2015}  In this case the leads frequently break, and even when successful the signal to noise is limited due to the small sample volume available.

An alternative approach is to perform optical detection through transparent diamond anvils. Optically detected magnetic resonance (ODMR) can be realized via negatively-charged nitrogen-vacancy (NV$^-$) centers in the diamond lattice.  NV$^-$ defects contain localized electronic states with a spin $S=1$ that can be probed via fluorescence spectroscopy. The spin-spin interactions lead to fine structure splitting in zero magnetic field, with a spin Hamiltonian:
\begin{equation}
\mathcal{H} = D\left(\hat{S}_z^2 - S(S+1)/3\right) + {E} \left(\hat{S}_x^2 - \hat{S}_{y}^2\right),
\end{equation}
where $D$ are $E$ are the zero-field splitting parameters.\cite{NVpressurePRL} $D$ and $E$ are determined by the unpaired electron spin density surrounding the defect, and depend on the details of the electronic wavefunction. For an isolated NV$^-$ center with $C_{3v}$ symmetry, $E$ vanishes, but in real systems strain transverse to the NV axis leads to a small splitting between the $S=\pm 1$ states in zero magnetic field.\cite{NVcenterGroupTheoryAnalysis} Single diamond crystals grown by chemical vapor deposition (CVD) with NV$^-$ centers are particularly important as magnetic field sensors, due to both the large gyromagnetic ratio of the electron spins  and the long spin coherence time $T_2 \sim 2 $ms  \cite{Balasubramanian2009} (up to 1s in specially-prepared crystals).\cite{BarGill2013}  NV centers have sensitivities on the order of nT/$\sqrt{\textrm{Hz}}$,\cite{Schoenfeld2011,Shin2012,NVcenterImaging2015} and have been used to detect nuclear spins outside of the diamond matrix.\cite{WratchrupScience2013,RugarNVScience2013}  As a result, NV$^-$ center magnetometry can offer an attractive alternative to conventional Faraday-induction-based magnetic resonance at high pressures in a DAC without the limited sensitivity and the technical difficulties of locating a detection coil within the sample space.

\begin{figure}[!tb]
\includegraphics[width=0.45\linewidth]{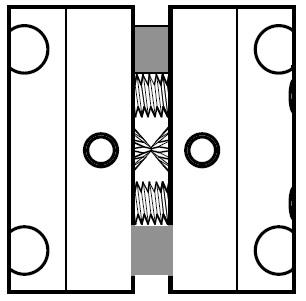}
\includegraphics[width=0.45\linewidth]{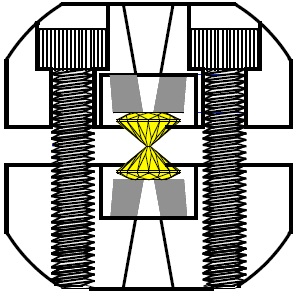}
\includegraphics[width=0.55\linewidth]{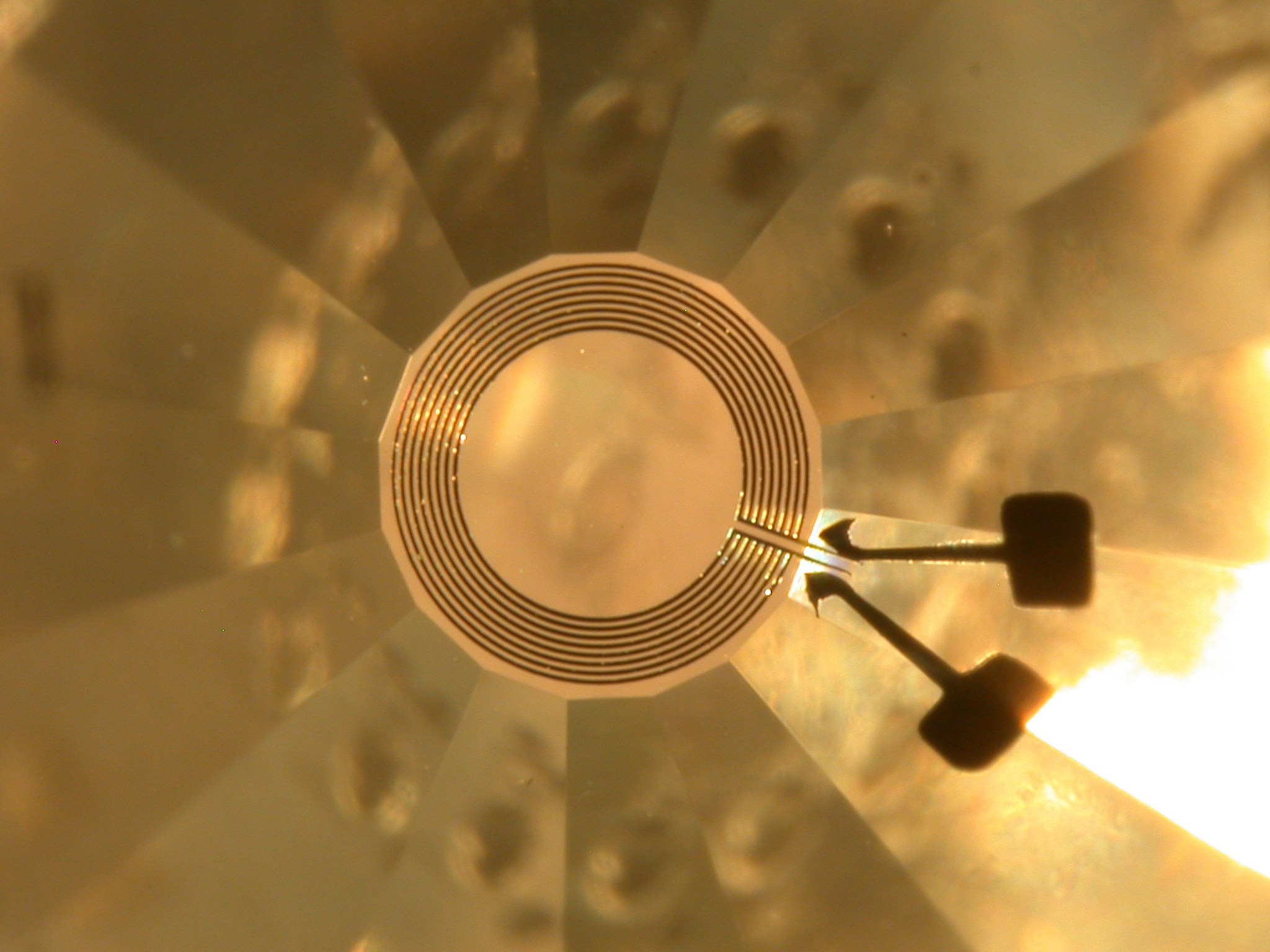}
\includegraphics[width=0.32\linewidth]{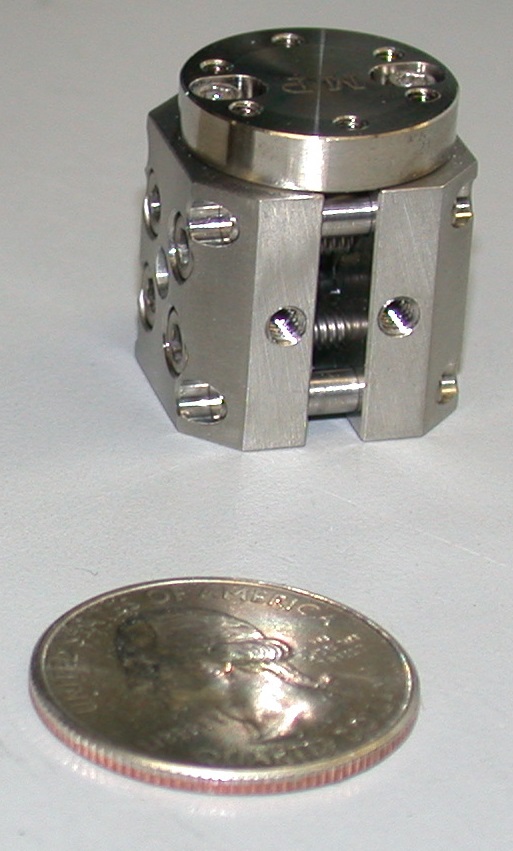}
\caption{\label{fig:DAC} Assembled DAC and designer anvil. The diameter of the culet is 1 mm. The tungsten microchannel consists of a series of eight concentric rings, each of width 5 $\mu$m. }
\end{figure}

ODMR has recently been performed on NV$^-$ centers up to 60 GPa in a diamond anvil cell.\cite{NVpressurePRL}   However, this required the use of a specialized non-conducting gasket in order to introduce the microwave radiation to the sample space between the anvils. The gasket consisted of an insulating matrix of boronitride powder mixed with epoxy, with an embedded Pt wire as a microwave antenna.  Here, we utilize a `designer' anvil with a metallic microchannel located within the culet of the anvil. This approach is superior because it enables us to lower the microwave power, reducing spurious heating effects, and does not require the use of non-conducting gasket materials.

\section{Pressure cell and Designer anvils}

We have developed a diamond anvil cell (DAC) for use in a small bore cryostat of inner diameter 25 mm, as shown in Fig. \ref{fig:DAC}. The cell is constructed from MP35N steel and is designed so that the pressure axis can be oriented perpendicular to the cryostat axis. Optical access to the sample is via a channel of length 7.3 mm to the anvil table, and opening angle of 20$^{\circ}$. The diamond anvils are type Ia gem-quality diamonds with 1mm diameter culets, and the anvils are secured by Stycast 1266 epoxy and aligned by eye under a microscope with the help of small set screws.  Pressure is applied by a set of four 4-40 steel bolts.

The gasket is manufactured by drilling a 300 $\mu$m diameter hole in MP35N steel  with a micro-EDM, followed by pre-indenting to a thickness of 100 $\mu$m.  The designer anvil was constructed by lithographically depositing tungsten metal with a pattern consisting of several concentric rings on the culet, with electrical contact pads located on the pavilion face.\cite{WierDACmicroheater,WeirSusceptibilityDACreview,WeirPRL}   After deposition, the a synthetic diamond layer grown by CVD was grown on the culet of thickness approximately 50$\mu$m.\cite{WeirDesignerDiamondLetter}  Electrical connections to the surface antenna were made with thin platinum wire and silver paint.  The antenna was then connected to an HP 8665A frequency source, which was itself controlled by a computer.

\section{Microdiamonds and Optical Design}

An ensemble of fluorescent microdiamonds  (Adamas nano) of average diameter $15\mu$m in powder form secured to the surface of the designer anvil within the sample space. In addition a small ruby chip of diameter 40 $\mu$m was located within the sample space in order to calibrate the pressure, \cite{RubyCalibration} and Daphne oil 7373 was used as a pressure medium.\cite{DaphneOilHighPressure}

\begin{figure}[h]
\includegraphics[width=\linewidth]{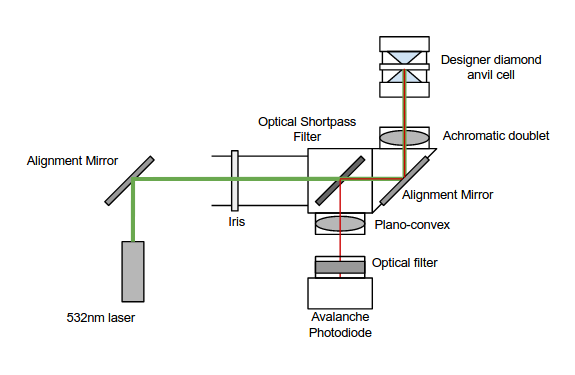}
\caption{\label{fig:setup} Optical layout for detecting the ODMR signal from the NV$^-$ centers. }
\end{figure}

The optical setup system consists of a 4.5 mW, 532 nm colimated laser diode source (Thorlabs CPS532) mounted on an optical kinematic mount (Thorlabs KM100) which is aligned into an optical cage cube system using a second kinematic mount with a 1" mirror (Thorlabs KM100-E02), as shown in Fig. \ref{fig:setup}.  The collimated beam passes through an optical shortpass filter (Edmund Optics 69-216) and is focused onto the NV$^-$ centers inside the sample space by using an achromatic doublet (Thorlabs AC254-030-A-ML) that focuses the beam through an optical access port in the DAC to a spot size of $\sim 10\mu$m.  The fluorescence is collected using the achromatic doublet, after which it reflects off the optical shortpass filter and is focused onto an avalanche photodiode (Thorlabs APD120A2/M) using a plano-convex lens (Thorlabs LA1131-A-ML).   ODMR spectra are collected by amplitude-modulating a continuous-wave microwave signal from the HP8665A with the reference of a SR510 lock-in amplifier, and measuring the output of the avalanche photodiode detector with the lock-in and a Keithley 192 Digital Voltmeter.\cite{ODMRnanodiamonds}
Control of the microwave source, lock-in and digital voltmeter for the experiments was achieved using Python PyVisa GPIB interfacing. The ruby fluorescence was measured and analyzed using an Ocean Optics HR4000 spectrometer.

\section{Results}

\begin{figure}[!tb]
\includegraphics[width=\linewidth]{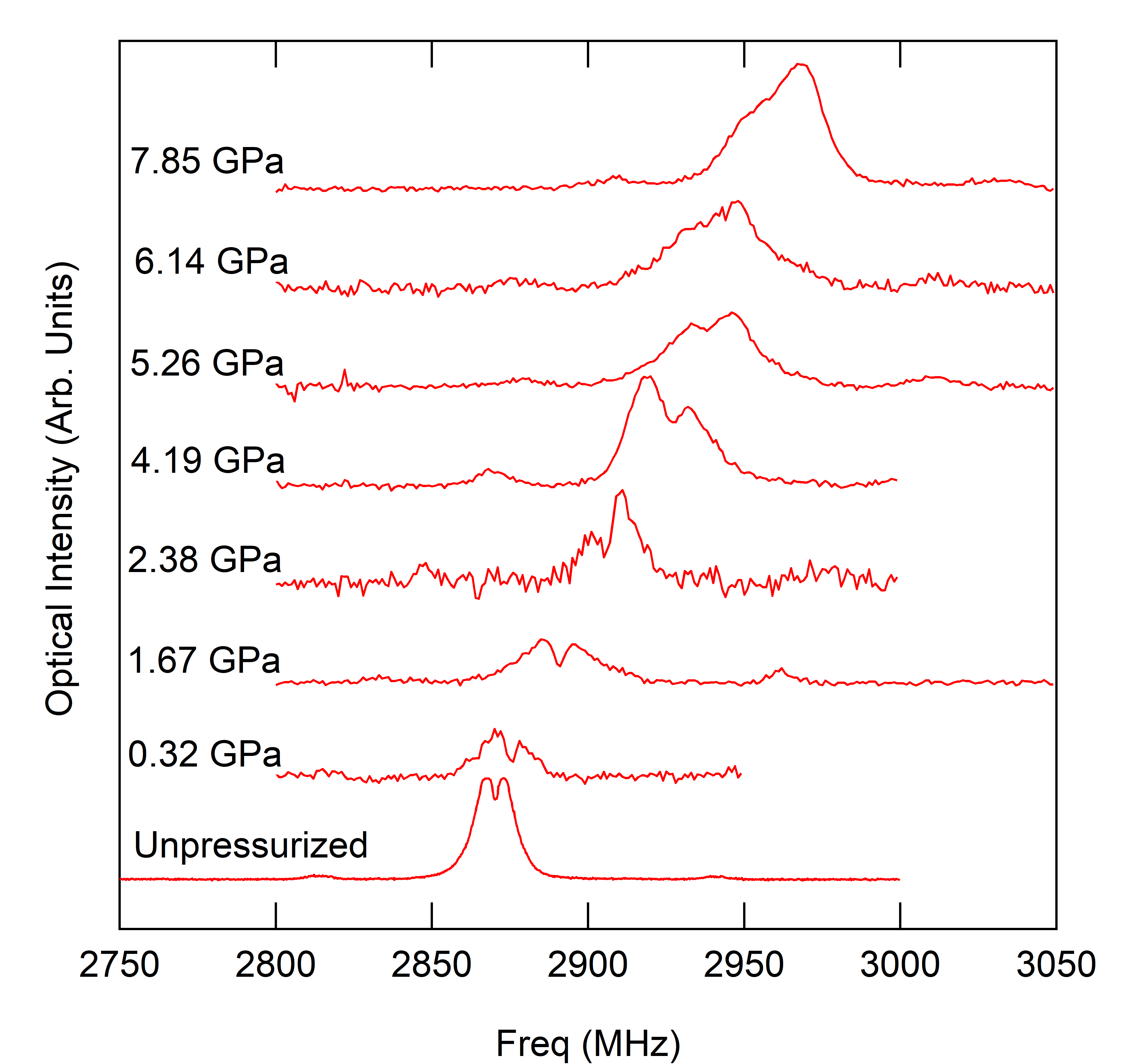}
\caption{\label{fig:waterfall} Zero-field ODMR spectra  of an ensemble of microcrystals at several pressures.}
\end{figure}

\begin{figure}[!tb]
\includegraphics[width=\linewidth]{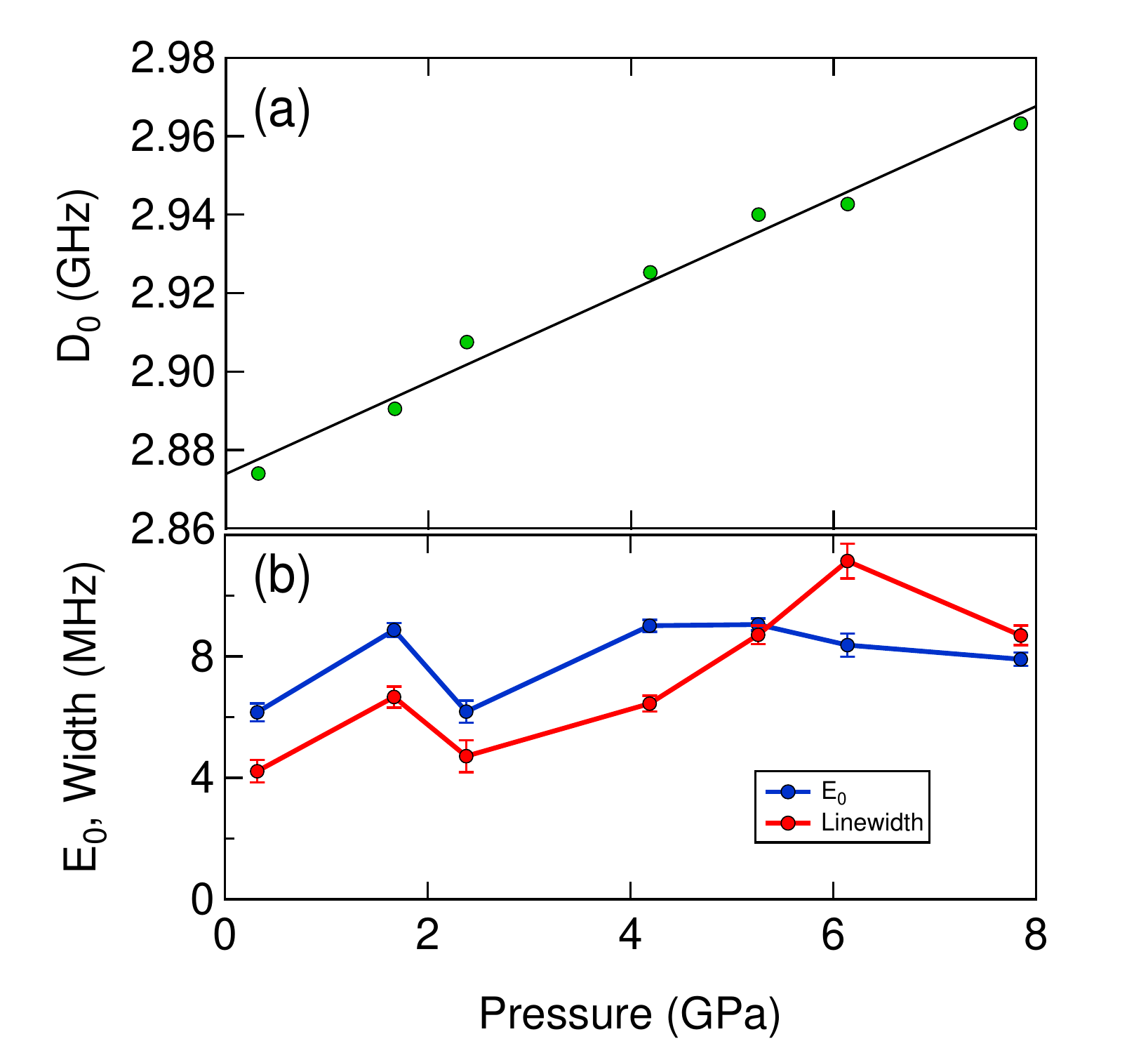}
\caption{\label{fig:parsVSpressure} Pressure dependence of  the zero-field splitting parameters, (a) $D$, and (b) $E_0$, as well as the linewidth. Error bars for (a) are the size of the points.}
\end{figure}

Figure \ref{fig:waterfall} shows the spectra of the NV$^-$ centers as a function of the microwave frequency at several different pressures.   We fit the spectra to the sum of two lorentzians, and Fig. \ref{fig:parsVSpressure} shows  the ZFS parameters $D$ and $E$.  The doublet feature arises due to the $E$ parameter.  $D$ varies linearly with pressure, with slope $dD/dP = 11.72 \pm 0.68$ MH/GPa. $D$ increases under pressure because the distance between the spins decreases  due to both the macroscopic compression of the diamond lattice, as well as local structural distortions at the NV defect site.\cite{NVpressureTheory}  The value we observe is slightly lower than reported previously, \cite{NVpressurePRL} but is close to the \textit{ab initio} calculation of 10.30 MHz/GPa.\cite{NVpressureTheory}  In the Doherty work, small chips taken from a type IIa CVD grown single crystal, whereas in this work commercial microdiamonds were measured . It is possible, therefore, that the difference in the pressure response could reflect differences in the sample preparation.  It is noteworthy that the anvils we used are type Ia, which nominally have a larger concentration of nitrogen impurities and hence a larger background fluorescence than type IIa anvils.  However, the contribution from any NV centers in the anvil is suppressed, since these are not located within the focal plan of the beam. As a result, the absorbtion and the fluorescence are both suppressed for NV centers outside of the microdiamonds.

Figure \ref{fig:parsVSpressure}(b) shows the linewidth and strain parameter $E$ as a function of pressure.  There is a slight increase in linewidth suggesting non-hydrostatic conditions within the cell.  A linear fit  yields a slope of 0.8 MHz/GPa, or a pressure variation $\delta P/P = 6\%$.     Daphne 7373 is expected to solidify at 2.2 GPa at room temperature, which roughly coincides with the increase observed in Fig. \ref{fig:parsVSpressure}.\cite{DaphneOilHighPressure}  $E$ does not show any significant variation with pressure, which likely reflects the fact that the local symmetry around the defect sites is not changing under pressure.\cite{NVcenterGroupTheoryAnalysis} Nevertheless, anisotropic strain  can contribute to linewidth broadening  because the shift will depend on the relative alignment of the strain tensor with any given NV$^-$ axis.  It is noteworthy that the relative intensities of the main doublet change with pressure, which suggests that inhomogeneous strain fields may be developing as the pressure medium solidifies.

\section{Conclusions}

Designer anvils offer a superior performance for ODMR of NV$^-$ centers in diamond anvil cells.  We have found that the the ZFS parameters for the NV$^-$ centers in microdiamonds exhibit a pressure dependence that agrees with \textit{ab initio} calculations. The linear response in the range of 0-8 GPa would enable the ODMR resonance to function as an effective manometer, eliminating the need to use a ruby chip and thus providing more space for the sample.  Furthermore, it may be possible to implant NV$^-$ centers directly on the culet through the CVD process for the designer anvil fabrication.  Although in this work an ensemble of microdiamonds was utilized, we have also observed ODMR from a single microdiamond crystal.  Single crystal ODMR would enable the application of magnetic fields to split the NV$^-$ resonance, which would enable vector magnetization measurements at high pressure.\cite{SteinartNVimaging} It would also enable the detection of nuclear spins on the microdiamond surface,\cite{WratchrupScience2013} opening a possible route to microscopic NMR measurements of superconducting sulfur hydride and metallic hydrogen.\cite{Drozdov2015,Dias715}

\begin{acknowledgments}
We thank  P. Klavins for assistance in the laboratory, and acknowledge discussions with X. Zhu. Work at UC Davis was supported by the NSF under Grant No.\ DMR-1506961, NSF grant PHY-1560482, and the NNSA under Grant No. DE-NA0002908. Work at LLNL was supported by UC Lab Fees Award \#12-LR-238151. Work at UC Berkeley was supported by the Director, Office of Science, Office of Basic Energy Sciences, Materials Sciences and Engineering Division, of the US Department of Energy under Contract No. DE-AC02-05CH11231. Lawrence Livermore National Laboratory is operated by Lawrence Livermore National Security, LLC, for the U.S. Department of Energy, National Nuclear Security Administration under Contract DE-AC52-07NA27344.  Work at the Univ. Alabama was supported by DOE-NNSA under Grant No. DE-NA0002928.

\end{acknowledgments}

\bibliography{CurroBibliography}

\begin{thebibliography}{27}%
\makeatletter
\providecommand \@ifxundefined [1]{%
 \@ifx{#1\undefined}
}%
\providecommand \@ifnum [1]{%
 \ifnum #1\expandafter \@firstoftwo
 \else \expandafter \@secondoftwo
 \fi
}%
\providecommand \@ifx [1]{%
 \ifx #1\expandafter \@firstoftwo
 \else \expandafter \@secondoftwo
 \fi
}%
\providecommand \natexlab [1]{#1}%
\providecommand \enquote  [1]{``#1''}%
\providecommand \bibnamefont  [1]{#1}%
\providecommand \bibfnamefont [1]{#1}%
\providecommand \citenamefont [1]{#1}%
\providecommand \href@noop [0]{\@secondoftwo}%
\providecommand \href [0]{\begingroup \@sanitize@url \@href}%
\providecommand \@href[1]{\@@startlink{#1}\@@href}%
\providecommand \@@href[1]{\endgroup#1\@@endlink}%
\providecommand \@sanitize@url [0]{\catcode `\\12\catcode `\$12\catcode
  `\&12\catcode `\#12\catcode `\^12\catcode `\_12\catcode `\%12\relax}%
\providecommand \@@startlink[1]{}%
\providecommand \@@endlink[0]{}%
\providecommand \url  [0]{\begingroup\@sanitize@url \@url }%
\providecommand \@url [1]{\endgroup\@href {#1}{\urlprefix }}%
\providecommand \urlprefix  [0]{URL }%
\providecommand \Eprint [0]{\href }%
\providecommand \doibase [0]{http://dx.doi.org/}%
\providecommand \selectlanguage [0]{\@gobble}%
\providecommand \bibinfo  [0]{\@secondoftwo}%
\providecommand \bibfield  [0]{\@secondoftwo}%
\providecommand \translation [1]{[#1]}%
\providecommand \BibitemOpen [0]{}%
\providecommand \bibitemStop [0]{}%
\providecommand \bibitemNoStop [0]{.\EOS\space}%
\providecommand \EOS [0]{\spacefactor3000\relax}%
\providecommand \BibitemShut  [1]{\csname bibitem#1\endcsname}%
\let\auto@bib@innerbib\@empty
\bibitem [{\citenamefont {Hegger}\ \emph {et~al.}(2000)\citenamefont {Hegger},
  \citenamefont {Petrovic}, \citenamefont {Moshopoulou}, \citenamefont
  {Hundley}, \citenamefont {Sarrao}, \citenamefont {Fisk},\ and\ \citenamefont
  {Thompson}}]{HeggerRh115discovery}%
  \BibitemOpen
  \bibfield  {author} {\bibinfo {author} {\bibfnamefont {H.}~\bibnamefont
  {Hegger}}, \bibinfo {author} {\bibfnamefont {C.}~\bibnamefont {Petrovic}},
  \bibinfo {author} {\bibfnamefont {E.~G.}\ \bibnamefont {Moshopoulou}},
  \bibinfo {author} {\bibfnamefont {M.~F.}\ \bibnamefont {Hundley}}, \bibinfo
  {author} {\bibfnamefont {J.~L.}\ \bibnamefont {Sarrao}}, \bibinfo {author}
  {\bibfnamefont {Z.}~\bibnamefont {Fisk}}, \ and\ \bibinfo {author}
  {\bibfnamefont {J.~D.}\ \bibnamefont {Thompson}},\ }\bibfield  {title}
  {\enquote {\bibinfo {title} {Pressure-induced superconductivity in quasi-{2D}
  {CeRhIn$_5$}},}\ }\href {\doibase 10.1103/PhysRevLett.84.4986} {\bibfield
  {journal} {\bibinfo  {journal} {Phys. Rev. Lett.}\ }\textbf {\bibinfo
  {volume} {84}},\ \bibinfo {pages} {4986--4989} (\bibinfo {year}
  {2000})}\BibitemShut {NoStop}%
\bibitem [{\citenamefont {Alireza}\ \emph {et~al.}(2009)\citenamefont
  {Alireza}, \citenamefont {Ko}, \citenamefont {Gillett}, \citenamefont
  {Petrone}, \citenamefont {Cole}, \citenamefont {Lonzarich},\ and\
  \citenamefont {Sebastian}}]{Cambridge122pressure2009}%
  \BibitemOpen
  \bibfield  {author} {\bibinfo {author} {\bibfnamefont {P.~L.}\ \bibnamefont
  {Alireza}}, \bibinfo {author} {\bibfnamefont {Y.~T.~C.}\ \bibnamefont {Ko}},
  \bibinfo {author} {\bibfnamefont {J.}~\bibnamefont {Gillett}}, \bibinfo
  {author} {\bibfnamefont {C.~M.}\ \bibnamefont {Petrone}}, \bibinfo {author}
  {\bibfnamefont {J.~M.}\ \bibnamefont {Cole}}, \bibinfo {author}
  {\bibfnamefont {G.~G.}\ \bibnamefont {Lonzarich}}, \ and\ \bibinfo {author}
  {\bibfnamefont {S.~E.}\ \bibnamefont {Sebastian}},\ }\bibfield  {title}
  {\enquote {\bibinfo {title} {Superconductivity up to 29 {K} in
  {SrFe$_2$As$_2$} and {BaFe$_2$As$_2$} at high pressures},}\ }\href
  {http://stacks.iop.org/0953-8984/21/i=1/a=012208} {\bibfield  {journal}
  {\bibinfo  {journal} {J. Phys.: Condens. Matter}\ }\textbf {\bibinfo {volume}
  {21}},\ \bibinfo {pages} {012208} (\bibinfo {year} {2009})}\BibitemShut
  {NoStop}%
\bibitem [{\citenamefont {Dias}\ and\ \citenamefont {Silvera}(2017)}]{Dias715}%
  \BibitemOpen
  \bibfield  {author} {\bibinfo {author} {\bibfnamefont {R.~P.}\ \bibnamefont
  {Dias}}\ and\ \bibinfo {author} {\bibfnamefont {I.~F.}\ \bibnamefont
  {Silvera}},\ }\bibfield  {title} {\enquote {\bibinfo {title} {Observation of
  the {Wigner-Huntington} transition to metallic hydrogen},}\ }\href {\doibase
  10.1126/science.aal1579} {\bibfield  {journal} {\bibinfo  {journal}
  {Science}\ }\textbf {\bibinfo {volume} {355}},\ \bibinfo {pages} {715--718}
  (\bibinfo {year} {2017})}\BibitemShut {NoStop}%
\bibitem [{\citenamefont {Kitagawa}\ \emph {et~al.}(2009)\citenamefont
  {Kitagawa}, \citenamefont {Katayama}, \citenamefont {Gotou}, \citenamefont
  {Yagi}, \citenamefont {Ohgushi}, \citenamefont {Matsumoto}, \citenamefont
  {Uwatoko},\ and\ \citenamefont {Takigawa}}]{TakigawaSr122pressure}%
  \BibitemOpen
  \bibfield  {author} {\bibinfo {author} {\bibfnamefont {K.}~\bibnamefont
  {Kitagawa}}, \bibinfo {author} {\bibfnamefont {N.}~\bibnamefont {Katayama}},
  \bibinfo {author} {\bibfnamefont {H.}~\bibnamefont {Gotou}}, \bibinfo
  {author} {\bibfnamefont {T.}~\bibnamefont {Yagi}}, \bibinfo {author}
  {\bibfnamefont {K.}~\bibnamefont {Ohgushi}}, \bibinfo {author} {\bibfnamefont
  {T.}~\bibnamefont {Matsumoto}}, \bibinfo {author} {\bibfnamefont
  {Y.}~\bibnamefont {Uwatoko}}, \ and\ \bibinfo {author} {\bibfnamefont
  {M.}~\bibnamefont {Takigawa}},\ }\bibfield  {title} {\enquote {\bibinfo
  {title} {Spontaneous formation of a superconducting and antiferromagnetic
  hybrid state in {SrF$e_{2}$As$_{2}$} under high pressure},}\ }\href {\doibase
  10.1103/PhysRevLett.103.257002} {\bibfield  {journal} {\bibinfo  {journal}
  {Phys. Rev. Lett.}\ }\textbf {\bibinfo {volume} {103}},\ \bibinfo {pages}
  {257002} (\bibinfo {year} {2009})}\BibitemShut {NoStop}%
\bibitem [{\citenamefont {Lin}\ \emph {et~al.}(2015)\citenamefont {Lin},
  \citenamefont {Shirer}, \citenamefont {Crocker}, \citenamefont {Dioguardi},
  \citenamefont {Lawson}, \citenamefont {Bush}, \citenamefont {Klavins},\ and\
  \citenamefont {Curro}}]{Lin2015}%
  \BibitemOpen
  \bibfield  {author} {\bibinfo {author} {\bibfnamefont {C.~H.}\ \bibnamefont
  {Lin}}, \bibinfo {author} {\bibfnamefont {K.~R.}\ \bibnamefont {Shirer}},
  \bibinfo {author} {\bibfnamefont {J.}~\bibnamefont {Crocker}}, \bibinfo
  {author} {\bibfnamefont {A.~P.}\ \bibnamefont {Dioguardi}}, \bibinfo {author}
  {\bibfnamefont {M.~M.}\ \bibnamefont {Lawson}}, \bibinfo {author}
  {\bibfnamefont {B.~T.}\ \bibnamefont {Bush}}, \bibinfo {author}
  {\bibfnamefont {P.}~\bibnamefont {Klavins}}, \ and\ \bibinfo {author}
  {\bibfnamefont {N.~J.}\ \bibnamefont {Curro}},\ }\bibfield  {title} {\enquote
  {\bibinfo {title} {Evolution of hyperfine parameters across a quantum
  critical point in {${\mathrm{CeRhIn}}_{5}$}},}\ }\href {\doibase
  10.1103/PhysRevB.92.155147} {\bibfield  {journal} {\bibinfo  {journal} {Phys.
  Rev. B}\ }\textbf {\bibinfo {volume} {92}},\ \bibinfo {pages} {155147}
  (\bibinfo {year} {2015})}\BibitemShut {NoStop}%
\bibitem [{\citenamefont {Kitagawa}\ \emph {et~al.}(2012)\citenamefont
  {Kitagawa}, \citenamefont {Matsubayashi}, \citenamefont {Gotou},
  \citenamefont {Matsumoto}, \citenamefont {Uwatoko}, \citenamefont {Yagi},\
  and\ \citenamefont {Takigawa}}]{KitagawaPressureCell}%
  \BibitemOpen
  \bibfield  {author} {\bibinfo {author} {\bibfnamefont {K.}~\bibnamefont
  {Kitagawa}}, \bibinfo {author} {\bibfnamefont {K.}~\bibnamefont
  {Matsubayashi}}, \bibinfo {author} {\bibfnamefont {H.}~\bibnamefont {Gotou}},
  \bibinfo {author} {\bibfnamefont {T.}~\bibnamefont {Matsumoto}}, \bibinfo
  {author} {\bibfnamefont {Y.}~\bibnamefont {Uwatoko}}, \bibinfo {author}
  {\bibfnamefont {T.}~\bibnamefont {Yagi}}, \ and\ \bibinfo {author}
  {\bibfnamefont {M.}~\bibnamefont {Takigawa}},\ }\bibfield  {title} {\enquote
  {\bibinfo {title} {10 {GPa}-class high-pressure {NMR} technique realized by
  the new cell with improved space efficiency},}\ }\href {\doibase
  10.4131/jshpreview.22.198} {\bibfield  {journal} {\bibinfo  {journal} {The
  Review of High Pressure Science and Technology}\ }\textbf {\bibinfo {volume}
  {22}},\ \bibinfo {pages} {198--205} (\bibinfo {year} {2012})}\BibitemShut
  {NoStop}%
\bibitem [{\citenamefont {Meier}, \citenamefont {Reichardt},\ and\
  \citenamefont {Haase}(2015)}]{HaaseJMR2015}%
  \BibitemOpen
  \bibfield  {author} {\bibinfo {author} {\bibfnamefont {T.}~\bibnamefont
  {Meier}}, \bibinfo {author} {\bibfnamefont {S.}~\bibnamefont {Reichardt}}, \
  and\ \bibinfo {author} {\bibfnamefont {J.}~\bibnamefont {Haase}},\ }\bibfield
   {title} {\enquote {\bibinfo {title} {High-sensitivity {NMR} beyond 200,000
  atmospheres of pressure},}\ }\href {\doibase
  http://dx.doi.org/10.1016/j.jmr.2015.05.007} {\bibfield  {journal} {\bibinfo
  {journal} {J. Magn. Reson.}\ }\textbf {\bibinfo {volume} {257}},\ \bibinfo
  {pages} {39 -- 44} (\bibinfo {year} {2015})}\BibitemShut {NoStop}%
\bibitem [{\citenamefont {Pravica}\ and\ \citenamefont
  {Silvera}(1998)}]{SilveraNMR}%
  \BibitemOpen
  \bibfield  {author} {\bibinfo {author} {\bibfnamefont {M.~G.}\ \bibnamefont
  {Pravica}}\ and\ \bibinfo {author} {\bibfnamefont {I.~F.}\ \bibnamefont
  {Silvera}},\ }\bibfield  {title} {\enquote {\bibinfo {title} {Nuclear
  magnetic resonance in a diamond anvil cell at very high pressures},}\ }\href
  {\doibase 10.1063/1.1148686} {\bibfield  {journal} {\bibinfo  {journal} {Rev.
  Sci. Instrum.}\ }\textbf {\bibinfo {volume} {69}},\ \bibinfo {pages} {479}
  (\bibinfo {year} {1998})}\BibitemShut {NoStop}%
\bibitem [{\citenamefont {Doherty}\ \emph {et~al.}(2014)\citenamefont
  {Doherty}, \citenamefont {Struzhkin}, \citenamefont {Simpson}, \citenamefont
  {McGuinness}, \citenamefont {Meng}, \citenamefont {Stacey}, \citenamefont
  {Karle}, \citenamefont {Hemley}, \citenamefont {Manson}, \citenamefont
  {Hollenberg},\ and\ \citenamefont {Prawer}}]{NVpressurePRL}%
  \BibitemOpen
  \bibfield  {author} {\bibinfo {author} {\bibfnamefont {M.~W.}\ \bibnamefont
  {Doherty}}, \bibinfo {author} {\bibfnamefont {V.~V.}\ \bibnamefont
  {Struzhkin}}, \bibinfo {author} {\bibfnamefont {D.~A.}\ \bibnamefont
  {Simpson}}, \bibinfo {author} {\bibfnamefont {L.~P.}\ \bibnamefont
  {McGuinness}}, \bibinfo {author} {\bibfnamefont {Y.}~\bibnamefont {Meng}},
  \bibinfo {author} {\bibfnamefont {A.}~\bibnamefont {Stacey}}, \bibinfo
  {author} {\bibfnamefont {T.~J.}\ \bibnamefont {Karle}}, \bibinfo {author}
  {\bibfnamefont {R.~J.}\ \bibnamefont {Hemley}}, \bibinfo {author}
  {\bibfnamefont {N.~B.}\ \bibnamefont {Manson}}, \bibinfo {author}
  {\bibfnamefont {L.~C.~L.}\ \bibnamefont {Hollenberg}}, \ and\ \bibinfo
  {author} {\bibfnamefont {S.}~\bibnamefont {Prawer}},\ }\bibfield  {title}
  {\enquote {\bibinfo {title} {Electronic properties and metrology applications
  of the diamond {${\mathrm{NV}}^{\ensuremath{-}}$} center under pressure},}\
  }\href {\doibase 10.1103/PhysRevLett.112.047601} {\bibfield  {journal}
  {\bibinfo  {journal} {Phys. Rev. Lett.}\ }\textbf {\bibinfo {volume} {112}},\
  \bibinfo {pages} {047601} (\bibinfo {year} {2014})}\BibitemShut {NoStop}%
\bibitem [{\citenamefont {Maze}\ \emph {et~al.}(2011)\citenamefont {Maze},
  \citenamefont {Gali}, \citenamefont {Togan}, \citenamefont {Chu},
  \citenamefont {Trifonov}, \citenamefont {Kaxiras},\ and\ \citenamefont
  {Lukin}}]{NVcenterGroupTheoryAnalysis}%
  \BibitemOpen
  \bibfield  {author} {\bibinfo {author} {\bibfnamefont {J.~R.}\ \bibnamefont
  {Maze}}, \bibinfo {author} {\bibfnamefont {A.}~\bibnamefont {Gali}}, \bibinfo
  {author} {\bibfnamefont {E.}~\bibnamefont {Togan}}, \bibinfo {author}
  {\bibfnamefont {Y.}~\bibnamefont {Chu}}, \bibinfo {author} {\bibfnamefont
  {A.}~\bibnamefont {Trifonov}}, \bibinfo {author} {\bibfnamefont
  {E.}~\bibnamefont {Kaxiras}}, \ and\ \bibinfo {author} {\bibfnamefont
  {M.~D.}\ \bibnamefont {Lukin}},\ }\bibfield  {title} {\enquote {\bibinfo
  {title} {Properties of nitrogen-vacancy centers in diamond: the group
  theoretic approach},}\ }\href
  {http://stacks.iop.org/1367-2630/13/i=2/a=025025} {\bibfield  {journal}
  {\bibinfo  {journal} {New J. Phys.}\ }\textbf {\bibinfo {volume} {13}},\
  \bibinfo {pages} {025025} (\bibinfo {year} {2011})}\BibitemShut {NoStop}%
\bibitem [{\citenamefont {Balasubramanian}\ \emph {et~al.}(2009)\citenamefont
  {Balasubramanian}, \citenamefont {Neumann}, \citenamefont {Twitchen},
  \citenamefont {Markham}, \citenamefont {Kolesov}, \citenamefont {Mizuochi},
  \citenamefont {Isoya}, \citenamefont {Achard}, \citenamefont {Beck},
  \citenamefont {Tissler}, \citenamefont {Jacques}, \citenamefont {Hemmer},
  \citenamefont {Jelezko},\ and\ \citenamefont
  {Wrachtrup}}]{Balasubramanian2009}%
  \BibitemOpen
  \bibfield  {author} {\bibinfo {author} {\bibfnamefont {G.}~\bibnamefont
  {Balasubramanian}}, \bibinfo {author} {\bibfnamefont {P.}~\bibnamefont
  {Neumann}}, \bibinfo {author} {\bibfnamefont {D.}~\bibnamefont {Twitchen}},
  \bibinfo {author} {\bibfnamefont {M.}~\bibnamefont {Markham}}, \bibinfo
  {author} {\bibfnamefont {R.}~\bibnamefont {Kolesov}}, \bibinfo {author}
  {\bibfnamefont {N.}~\bibnamefont {Mizuochi}}, \bibinfo {author}
  {\bibfnamefont {J.}~\bibnamefont {Isoya}}, \bibinfo {author} {\bibfnamefont
  {J.}~\bibnamefont {Achard}}, \bibinfo {author} {\bibfnamefont
  {J.}~\bibnamefont {Beck}}, \bibinfo {author} {\bibfnamefont {J.}~\bibnamefont
  {Tissler}}, \bibinfo {author} {\bibfnamefont {V.}~\bibnamefont {Jacques}},
  \bibinfo {author} {\bibfnamefont {P.~R.}\ \bibnamefont {Hemmer}}, \bibinfo
  {author} {\bibfnamefont {F.}~\bibnamefont {Jelezko}}, \ and\ \bibinfo
  {author} {\bibfnamefont {J.}~\bibnamefont {Wrachtrup}},\ }\bibfield  {title}
  {\enquote {\bibinfo {title} {Ultralong spin coherence time in isotopically
  engineered diamond},}\ }\href {http://dx.doi.org/10.1038/nmat2420} {\bibfield
   {journal} {\bibinfo  {journal} {Nat. Mater.}\ }\textbf {\bibinfo {volume}
  {8}},\ \bibinfo {pages} {383--387} (\bibinfo {year} {2009})}\BibitemShut
  {NoStop}%
\bibitem [{\citenamefont {Bar-Gill}\ \emph {et~al.}(2013)\citenamefont
  {Bar-Gill}, \citenamefont {Pham}, \citenamefont {Jarmola}, \citenamefont
  {Budker},\ and\ \citenamefont {Walsworth}}]{BarGill2013}%
  \BibitemOpen
  \bibfield  {author} {\bibinfo {author} {\bibfnamefont {N.}~\bibnamefont
  {Bar-Gill}}, \bibinfo {author} {\bibfnamefont {L.~M.}\ \bibnamefont {Pham}},
  \bibinfo {author} {\bibfnamefont {A.}~\bibnamefont {Jarmola}}, \bibinfo
  {author} {\bibfnamefont {D.}~\bibnamefont {Budker}}, \ and\ \bibinfo {author}
  {\bibfnamefont {R.~L.}\ \bibnamefont {Walsworth}},\ }\bibfield  {title}
  {\enquote {\bibinfo {title} {Solid-state electronic spin coherence time
  approaching one second},}\ }\href {http://dx.doi.org/10.1038/ncomms2771} {\
  \textbf {\bibinfo {volume} {4}},\ \bibinfo {pages} {1743} (\bibinfo {year}
  {2013})}\BibitemShut {NoStop}%
\bibitem [{\citenamefont {Schoenfeld}\ and\ \citenamefont
  {Harneit}(2011)}]{Schoenfeld2011}%
  \BibitemOpen
  \bibfield  {author} {\bibinfo {author} {\bibfnamefont {R.~S.}\ \bibnamefont
  {Schoenfeld}}\ and\ \bibinfo {author} {\bibfnamefont {W.}~\bibnamefont
  {Harneit}},\ }\bibfield  {title} {\enquote {\bibinfo {title} {Real time
  magnetic field sensing and imaging using a single spin in diamond},}\ }\href
  {\doibase 10.1103/physrevlett.106.030802} {\bibfield  {journal} {\bibinfo
  {journal} {Phys. Rev. Lett.}\ }\textbf {\bibinfo {volume} {106}},\ \bibinfo
  {pages} {030802} (\bibinfo {year} {2011})}\BibitemShut {NoStop}%
\bibitem [{\citenamefont {Shin}\ \emph {et~al.}(2012)\citenamefont {Shin},
  \citenamefont {Avalos}, \citenamefont {Butler}, \citenamefont {Trease},
  \citenamefont {Seltzer}, \citenamefont {Peter~Mustonen}, \citenamefont
  {Kennedy}, \citenamefont {Acosta}, \citenamefont {Budker}, \citenamefont
  {Pines},\ and\ \citenamefont {et~al.}}]{Shin2012}%
  \BibitemOpen
  \bibfield  {author} {\bibinfo {author} {\bibfnamefont {C.~S.}\ \bibnamefont
  {Shin}}, \bibinfo {author} {\bibfnamefont {C.~E.}\ \bibnamefont {Avalos}},
  \bibinfo {author} {\bibfnamefont {M.~C.}\ \bibnamefont {Butler}}, \bibinfo
  {author} {\bibfnamefont {D.~R.}\ \bibnamefont {Trease}}, \bibinfo {author}
  {\bibfnamefont {S.~J.}\ \bibnamefont {Seltzer}}, \bibinfo {author}
  {\bibfnamefont {J.}~\bibnamefont {Peter~Mustonen}}, \bibinfo {author}
  {\bibfnamefont {D.~J.}\ \bibnamefont {Kennedy}}, \bibinfo {author}
  {\bibfnamefont {V.~M.}\ \bibnamefont {Acosta}}, \bibinfo {author}
  {\bibfnamefont {D.}~\bibnamefont {Budker}}, \bibinfo {author} {\bibfnamefont
  {A.}~\bibnamefont {Pines}}, \ and\ \bibinfo {author} {\bibnamefont
  {et~al.}},\ }\bibfield  {title} {\enquote {\bibinfo {title} {Room-temperature
  operation of a radiofrequency diamond magnetometer near the shot-noise
  limit},}\ }\href {\doibase 10.1063/1.4771924} {\bibfield  {journal} {\bibinfo
   {journal} {J. Appl. Phys.}\ }\textbf {\bibinfo {volume} {112}},\ \bibinfo
  {pages} {124519} (\bibinfo {year} {2012})}\BibitemShut {NoStop}%
\bibitem [{\citenamefont {Ajoy}\ \emph {et~al.}(2015)\citenamefont {Ajoy},
  \citenamefont {Bissbort}, \citenamefont {Lukin}, \citenamefont {Walsworth},\
  and\ \citenamefont {Cappellaro}}]{NVcenterImaging2015}%
  \BibitemOpen
  \bibfield  {author} {\bibinfo {author} {\bibfnamefont {A.}~\bibnamefont
  {Ajoy}}, \bibinfo {author} {\bibfnamefont {U.}~\bibnamefont {Bissbort}},
  \bibinfo {author} {\bibfnamefont {M.~D.}\ \bibnamefont {Lukin}}, \bibinfo
  {author} {\bibfnamefont {R.~L.}\ \bibnamefont {Walsworth}}, \ and\ \bibinfo
  {author} {\bibfnamefont {P.}~\bibnamefont {Cappellaro}},\ }\bibfield  {title}
  {\enquote {\bibinfo {title} {Atomic-scale nuclear spin imaging using
  quantum-assisted sensors in diamond},}\ }\href {\doibase
  10.1103/PhysRevX.5.011001} {\bibfield  {journal} {\bibinfo  {journal} {Phys.
  Rev. X}\ }\textbf {\bibinfo {volume} {5}},\ \bibinfo {pages} {011001}
  (\bibinfo {year} {2015})}\BibitemShut {NoStop}%
\bibitem [{\citenamefont {Staudacher}\ \emph {et~al.}(2013)\citenamefont
  {Staudacher}, \citenamefont {Shi}, \citenamefont {Pezzagna}, \citenamefont
  {Meijer}, \citenamefont {Du}, \citenamefont {Meriles}, \citenamefont
  {Reinhard},\ and\ \citenamefont {Wrachtrup}}]{WratchrupScience2013}%
  \BibitemOpen
  \bibfield  {author} {\bibinfo {author} {\bibfnamefont {T.}~\bibnamefont
  {Staudacher}}, \bibinfo {author} {\bibfnamefont {F.}~\bibnamefont {Shi}},
  \bibinfo {author} {\bibfnamefont {S.}~\bibnamefont {Pezzagna}}, \bibinfo
  {author} {\bibfnamefont {J.}~\bibnamefont {Meijer}}, \bibinfo {author}
  {\bibfnamefont {J.}~\bibnamefont {Du}}, \bibinfo {author} {\bibfnamefont
  {C.~A.}\ \bibnamefont {Meriles}}, \bibinfo {author} {\bibfnamefont
  {F.}~\bibnamefont {Reinhard}}, \ and\ \bibinfo {author} {\bibfnamefont
  {J.}~\bibnamefont {Wrachtrup}},\ }\bibfield  {title} {\enquote {\bibinfo
  {title} {Nuclear magnetic resonance spectroscopy on a (5-nanometer)$^3$
  sample volume},}\ }\href {\doibase 10.1126/science.1231675} {\bibfield
  {journal} {\bibinfo  {journal} {Science}\ }\textbf {\bibinfo {volume}
  {339}},\ \bibinfo {pages} {561--563} (\bibinfo {year} {2013})}\BibitemShut
  {NoStop}%
\bibitem [{\citenamefont {Mamin}\ \emph {et~al.}(2013)\citenamefont {Mamin},
  \citenamefont {Kim}, \citenamefont {Sherwood}, \citenamefont {Rettner},
  \citenamefont {Ohno}, \citenamefont {Awschalom},\ and\ \citenamefont
  {Rugar}}]{RugarNVScience2013}%
  \BibitemOpen
  \bibfield  {author} {\bibinfo {author} {\bibfnamefont {H.~J.}\ \bibnamefont
  {Mamin}}, \bibinfo {author} {\bibfnamefont {M.}~\bibnamefont {Kim}}, \bibinfo
  {author} {\bibfnamefont {M.~H.}\ \bibnamefont {Sherwood}}, \bibinfo {author}
  {\bibfnamefont {C.~T.}\ \bibnamefont {Rettner}}, \bibinfo {author}
  {\bibfnamefont {K.}~\bibnamefont {Ohno}}, \bibinfo {author} {\bibfnamefont
  {D.~D.}\ \bibnamefont {Awschalom}}, \ and\ \bibinfo {author} {\bibfnamefont
  {D.}~\bibnamefont {Rugar}},\ }\bibfield  {title} {\enquote {\bibinfo {title}
  {Nanoscale nuclear magnetic resonance with a nitrogen-vacancy spin sensor},}\
  }\href {\doibase 10.1126/science.1231540} {\bibfield  {journal} {\bibinfo
  {journal} {Science}\ }\textbf {\bibinfo {volume} {339}},\ \bibinfo {pages}
  {557--560} (\bibinfo {year} {2013})}\BibitemShut {NoStop}%
\bibitem [{\citenamefont {Weir}\ \emph {et~al.}(2009)\citenamefont {Weir},
  \citenamefont {Jackson}, \citenamefont {Falabella}, \citenamefont
  {Samudrala},\ and\ \citenamefont {Vohra}}]{WierDACmicroheater}%
  \BibitemOpen
  \bibfield  {author} {\bibinfo {author} {\bibfnamefont {S.~T.}\ \bibnamefont
  {Weir}}, \bibinfo {author} {\bibfnamefont {D.~D.}\ \bibnamefont {Jackson}},
  \bibinfo {author} {\bibfnamefont {S.}~\bibnamefont {Falabella}}, \bibinfo
  {author} {\bibfnamefont {G.}~\bibnamefont {Samudrala}}, \ and\ \bibinfo
  {author} {\bibfnamefont {Y.~K.}\ \bibnamefont {Vohra}},\ }\bibfield  {title}
  {\enquote {\bibinfo {title} {An electrical microheater technique for
  high-pressure and high-temperature diamond anvil cell experiments},}\ }\href
  {\doibase 10.1063/1.3069286} {\bibfield  {journal} {\bibinfo  {journal} {Rev.
  Sci. Instrum.}\ }\textbf {\bibinfo {volume} {80}},\ \bibinfo {pages}
  {013905­013910} (\bibinfo {year} {2009})}\BibitemShut {NoStop}%
\bibitem [{\citenamefont {Jackson}\ \emph {et~al.}(2003)\citenamefont
  {Jackson}, \citenamefont {Aracne-Ruddle}, \citenamefont {Malba},
  \citenamefont {Weir}, \citenamefont {Catledge},\ and\ \citenamefont
  {Vohra}}]{WeirSusceptibilityDACreview}%
  \BibitemOpen
  \bibfield  {author} {\bibinfo {author} {\bibfnamefont {D.~D.}\ \bibnamefont
  {Jackson}}, \bibinfo {author} {\bibfnamefont {C.}~\bibnamefont
  {Aracne-Ruddle}}, \bibinfo {author} {\bibfnamefont {V.}~\bibnamefont
  {Malba}}, \bibinfo {author} {\bibfnamefont {S.~T.}\ \bibnamefont {Weir}},
  \bibinfo {author} {\bibfnamefont {S.~A.}\ \bibnamefont {Catledge}}, \ and\
  \bibinfo {author} {\bibfnamefont {Y.~K.}\ \bibnamefont {Vohra}},\ }\bibfield
  {title} {\enquote {\bibinfo {title} {Magnetic susceptibility measurements at
  high pressure using designer diamond anvils},}\ }\href {\doibase
  10.1063/1.1544084} {\bibfield  {journal} {\bibinfo  {journal} {Rev. Sci.
  Instrum.}\ }\textbf {\bibinfo {volume} {74}},\ \bibinfo {pages} {2467}
  (\bibinfo {year} {2003})}\BibitemShut {NoStop}%
\bibitem [{\citenamefont {Patterson}\ \emph {et~al.}(2000)\citenamefont
  {Patterson}, \citenamefont {Catledge}, \citenamefont {Vohra}, \citenamefont
  {Akella},\ and\ \citenamefont {Weir}}]{WeirPRL}%
  \BibitemOpen
  \bibfield  {author} {\bibinfo {author} {\bibfnamefont {J.~R.}\ \bibnamefont
  {Patterson}}, \bibinfo {author} {\bibfnamefont {S.~A.}\ \bibnamefont
  {Catledge}}, \bibinfo {author} {\bibfnamefont {Y.~K.}\ \bibnamefont {Vohra}},
  \bibinfo {author} {\bibfnamefont {J.}~\bibnamefont {Akella}}, \ and\ \bibinfo
  {author} {\bibfnamefont {S.~T.}\ \bibnamefont {Weir}},\ }\bibfield  {title}
  {\enquote {\bibinfo {title} {Electrical and mechanical properties of
  ${C}_{70}$ fullerene and graphite under high pressures studied using designer
  diamond anvils},}\ }\href {\doibase 10.1103/PhysRevLett.85.5364} {\bibfield
  {journal} {\bibinfo  {journal} {Phys. Rev. Lett.}\ }\textbf {\bibinfo
  {volume} {85}},\ \bibinfo {pages} {5364--5367} (\bibinfo {year}
  {2000})}\BibitemShut {NoStop}%
\bibitem [{\citenamefont {Weir}\ \emph {et~al.}(2000)\citenamefont {Weir},
  \citenamefont {Akella}, \citenamefont {Aracne-Ruddle}, \citenamefont
  {Vohra},\ and\ \citenamefont {Catledge}}]{WeirDesignerDiamondLetter}%
  \BibitemOpen
  \bibfield  {author} {\bibinfo {author} {\bibfnamefont {S.~T.}\ \bibnamefont
  {Weir}}, \bibinfo {author} {\bibfnamefont {J.}~\bibnamefont {Akella}},
  \bibinfo {author} {\bibfnamefont {C.}~\bibnamefont {Aracne-Ruddle}}, \bibinfo
  {author} {\bibfnamefont {Y.~K.}\ \bibnamefont {Vohra}}, \ and\ \bibinfo
  {author} {\bibfnamefont {S.~A.}\ \bibnamefont {Catledge}},\ }\bibfield
  {title} {\enquote {\bibinfo {title} {Epitaxial diamond encapsulation of metal
  microprobes for high pressure experiments},}\ }\href {\doibase
  10.1063/1.1326838} {\bibfield  {journal} {\bibinfo  {journal} {Appl. Phys.
  Lett.}\ }\textbf {\bibinfo {volume} {77}},\ \bibinfo {pages} {3400} (\bibinfo
  {year} {2000})}\BibitemShut {NoStop}%
\bibitem [{\citenamefont {Mao}, \citenamefont {Xu},\ and\ \citenamefont
  {Bell}(1986)}]{RubyCalibration}%
  \BibitemOpen
  \bibfield  {author} {\bibinfo {author} {\bibfnamefont {H.~K.}\ \bibnamefont
  {Mao}}, \bibinfo {author} {\bibfnamefont {J.}~\bibnamefont {Xu}}, \ and\
  \bibinfo {author} {\bibfnamefont {P.~M.}\ \bibnamefont {Bell}},\ }\bibfield
  {title} {\enquote {\bibinfo {title} {Calibration of the ruby pressure gauge
  to 800 kbar under quasi-hydrostatic conditions},}\ }\href {\doibase
  10.1029/JB091iB05p04673} {\bibfield  {journal} {\bibinfo  {journal} {Journal
  of Geophysical Research: Solid Earth}\ }\textbf {\bibinfo {volume} {91}},\
  \bibinfo {pages} {4673--4676} (\bibinfo {year} {1986})}\BibitemShut {NoStop}%
\bibitem [{\citenamefont {Yokogawa}\ \emph {et~al.}(2007)\citenamefont
  {Yokogawa}, \citenamefont {Murata}, \citenamefont {Yoshino},\ and\
  \citenamefont {Aoyama}}]{DaphneOilHighPressure}%
  \BibitemOpen
  \bibfield  {author} {\bibinfo {author} {\bibfnamefont {K.}~\bibnamefont
  {Yokogawa}}, \bibinfo {author} {\bibfnamefont {K.}~\bibnamefont {Murata}},
  \bibinfo {author} {\bibfnamefont {H.}~\bibnamefont {Yoshino}}, \ and\
  \bibinfo {author} {\bibfnamefont {S.}~\bibnamefont {Aoyama}},\ }\bibfield
  {title} {\enquote {\bibinfo {title} {Solidification of high-pressure medium
  {Daphne} 7373},}\ }\href {http://stacks.iop.org/1347-4065/46/i=6R/a=3636}
  {\bibfield  {journal} {\bibinfo  {journal} {Jpn. J. Appl. Phys.}\ }\textbf
  {\bibinfo {volume} {46}},\ \bibinfo {pages} {3636} (\bibinfo {year}
  {2007})}\BibitemShut {NoStop}%
\bibitem [{\citenamefont {Jeong}\ \emph {et~al.}(2017)\citenamefont {Jeong},
  \citenamefont {Parker}, \citenamefont {Page}, \citenamefont {Pines},
  \citenamefont {Vassiliou},\ and\ \citenamefont {King}}]{ODMRnanodiamonds}%
  \BibitemOpen
  \bibfield  {author} {\bibinfo {author} {\bibfnamefont {K.}~\bibnamefont
  {Jeong}}, \bibinfo {author} {\bibfnamefont {A.~J.}\ \bibnamefont {Parker}},
  \bibinfo {author} {\bibfnamefont {R.~H.}\ \bibnamefont {Page}}, \bibinfo
  {author} {\bibfnamefont {A.}~\bibnamefont {Pines}}, \bibinfo {author}
  {\bibfnamefont {C.~C.}\ \bibnamefont {Vassiliou}}, \ and\ \bibinfo {author}
  {\bibfnamefont {J.~P.}\ \bibnamefont {King}},\ }\bibfield  {title} {\enquote
  {\bibinfo {title} {Understanding the magnetic resonance spectrum of nitrogen
  vacancy centers in an ensemble of randomly-oriented nanodiamonds},}\ }\href
  {\doibase 10.1021/acs.jpcc.7b07247} {\bibfield  {journal} {\bibinfo
  {journal} {The Journal of Physical Chemistry C}\ } (\bibinfo {year} {2017}),\
  10.1021/acs.jpcc.7b07247}\BibitemShut {NoStop}%
\bibitem [{\citenamefont {Iv\'ady}\ \emph {et~al.}(2014)\citenamefont
  {Iv\'ady}, \citenamefont {Simon}, \citenamefont {Maze}, \citenamefont
  {Abrikosov},\ and\ \citenamefont {Gali}}]{NVpressureTheory}%
  \BibitemOpen
  \bibfield  {author} {\bibinfo {author} {\bibfnamefont {V.}~\bibnamefont
  {Iv\'ady}}, \bibinfo {author} {\bibfnamefont {T.}~\bibnamefont {Simon}},
  \bibinfo {author} {\bibfnamefont {J.~R.}\ \bibnamefont {Maze}}, \bibinfo
  {author} {\bibfnamefont {I.~A.}\ \bibnamefont {Abrikosov}}, \ and\ \bibinfo
  {author} {\bibfnamefont {A.}~\bibnamefont {Gali}},\ }\bibfield  {title}
  {\enquote {\bibinfo {title} {Pressure and temperature dependence of the
  zero-field splitting in the ground state of {NV} centers in diamond: A
  first-principles study},}\ }\href {\doibase 10.1103/PhysRevB.90.235205}
  {\bibfield  {journal} {\bibinfo  {journal} {Phys. Rev. B}\ }\textbf {\bibinfo
  {volume} {90}},\ \bibinfo {pages} {235205} (\bibinfo {year}
  {2014})}\BibitemShut {NoStop}%
\bibitem [{\citenamefont {Steinert}\ \emph {et~al.}(2010)\citenamefont
  {Steinert}, \citenamefont {Dolde}, \citenamefont {Neumann}, \citenamefont
  {Aird}, \citenamefont {Naydenov}, \citenamefont {Balasubramanian},
  \citenamefont {Jelezko},\ and\ \citenamefont
  {Wrachtrup}}]{SteinartNVimaging}%
  \BibitemOpen
  \bibfield  {author} {\bibinfo {author} {\bibfnamefont {S.}~\bibnamefont
  {Steinert}}, \bibinfo {author} {\bibfnamefont {F.}~\bibnamefont {Dolde}},
  \bibinfo {author} {\bibfnamefont {P.}~\bibnamefont {Neumann}}, \bibinfo
  {author} {\bibfnamefont {A.}~\bibnamefont {Aird}}, \bibinfo {author}
  {\bibfnamefont {B.}~\bibnamefont {Naydenov}}, \bibinfo {author}
  {\bibfnamefont {G.}~\bibnamefont {Balasubramanian}}, \bibinfo {author}
  {\bibfnamefont {F.}~\bibnamefont {Jelezko}}, \ and\ \bibinfo {author}
  {\bibfnamefont {J.}~\bibnamefont {Wrachtrup}},\ }\bibfield  {title} {\enquote
  {\bibinfo {title} {High sensitivity magnetic imaging using an array of spins
  in diamond},}\ }\href {\doibase 10.1063/1.3385689} {\bibfield  {journal}
  {\bibinfo  {journal} {Rev. Sci. Instrum.}\ }\textbf {\bibinfo {volume}
  {81}},\ \bibinfo {pages} {043705} (\bibinfo {year} {2010})}\BibitemShut
  {NoStop}%
\bibitem [{\citenamefont {Drozdov}\ \emph {et~al.}(2015)\citenamefont
  {Drozdov}, \citenamefont {Eremets}, \citenamefont {Troyan}, \citenamefont
  {Ksenofontov},\ and\ \citenamefont {Shylin}}]{Drozdov2015}%
  \BibitemOpen
  \bibfield  {author} {\bibinfo {author} {\bibfnamefont {A.~P.}\ \bibnamefont
  {Drozdov}}, \bibinfo {author} {\bibfnamefont {M.~I.}\ \bibnamefont
  {Eremets}}, \bibinfo {author} {\bibfnamefont {I.~A.}\ \bibnamefont {Troyan}},
  \bibinfo {author} {\bibfnamefont {V.}~\bibnamefont {Ksenofontov}}, \ and\
  \bibinfo {author} {\bibfnamefont {S.~I.}\ \bibnamefont {Shylin}},\ }\bibfield
   {title} {\enquote {\bibinfo {title} {Conventional superconductivity at 203
  kelvin at high pressures in the sulfur hydride system},}\ }\href
  {http://dx.doi.org/10.1038/nature14964} {\bibfield  {journal} {\bibinfo
  {journal} {Nature}\ }\textbf {\bibinfo {volume} {525}},\ \bibinfo {pages}
  {73--76} (\bibinfo {year} {2015})}\BibitemShut {NoStop}%
\end{thebibliography}%

\end{document}